\documentstyle[preprint,aps]{revtex}
\input epsf.tex
\def\DESepsf(#1 width #2){\epsfxsize=#2 \epsfbox{#1}}
\begin{document}

\preprint{\vbox{\hbox{UMD-PP-97-102
   }\hbox{OITS-625}\hbox{OSURN-324}\hbox{}}}
\draft
\title {\Large \bf Sparticle Spectroscopy and Phenomenology  \\ in a New Class 
of Gauge Mediated \\Supersymmetry Breaking Models }
\author{\bf Z. Chacko$^{1}$, B. Dutta$^{2}$, R. N. Mohapatra$^{1}$  and S.
Nandi$^{3}$}
\address{ $^{1}$ Department of Physics, University of Maryland, College Park,
 MD-20742\\
$^{2}$ Institute of Theoretical Science, University of Oregon, Eugene, OR 97403\\
$^{3}$ Department of Physics, Oklahoma State University, Stillwater, OK-74078 }
\date{April, 1997)\\\noindent (To appear in Phys. Rev. D}
\maketitle
\begin{abstract} Recently, a proposal (by R.N.M. and S.N.) was  made for a new
class of gauge  mediated supersymmetry breaking (GMSB) models  where the
standard model gauge group is embedded into the gauge group
$SU(2)_L\times U(1)_{I_{3R}}\times U(1)_{B-L}$ (or $SU(2)_L\times SU(2)_R\times
U(1)_{B-L}$) at the supersymmetry breaking scale $\Lambda$. Supersymmetry
breaking is transmitted to the visible sector via the same fields that are
responsible for gauge symmetry breaking rather than by vector-like quarks and
leptons as in the conventional GMSB models. These models have a number of
attractive properties such as  exact R-parity conservation, non-vanishing 
neutrino masses and a solution to the SUSYCP (and strong CP) problem. In this
paper, we present the detailed sparticle spectroscopy and phenomenological
implications of the various models of this class that embody the general spirit
of our previous  work but use a larger variety of messenger fields. A distinct
characteristic of this class of models is that unlike the conventional GMSB
ones, the lightest neutralino is always the NLSP leading to photonic events in
the colliders.    
\end{abstract}
\newpage 

\section{Introduction}
                                                        
There are many compelling reasons to believe  that nature is supersymmetric at
short distances. Since the observed spectrum of fermions and gauge bosons does
not exhibit any trace of supersymmetry, it must be broken at a scale around or
higher than 100 GeV. The general  procedure followed in building realistic
models with broken supersymmetry is to assume that supersymmetry breaking takes
place in a hidden sector which is completely separate from the visible sector
of the standard model and  have this effect transmitted to the quarks and
leptons via a messenger sector. Different classes of supersymmetric models can
be distinguished by the way the messengers transmit supersymmetry breaking from
the hidden sector to the  visible one. A scenario which has become very popular
in  the last two years is one where the messenger fields are replicas of the
known quarks and leptons except that they are heavier and they come in vector
like pairs\cite{dine} and the  standard model gauge interactions of the
messenger fields are the ones that transmit supersymmetry breaking to the
visible sector. The advantage of these models is that they lead to degenerate
squark and slepton masses at the scale $\Lambda$ which then provides a natural
solution to the flavor changing neutral current (FCNC) problem of the low energy
supersymmetry models. This makes the models phenomenologically very attractive.
Secondly these models are extremely predictive\cite{GMSB} so that one can have a
genuine hope that they can be experimentally tested in the not too distant
future.

There are however several drawbacks of these models: (i) one needs to put in
extra vectorlike quarks and leptons whose sole purpose is to transmit the
supersymmetry breaking from the hidden to the visible sector; (ii) it has been
argued\cite{randall} that in explicit models of supersymmetry breaking in the
hidden sector, the lowest vacuum breaks color and only the false vacuum has the
desirable properties; (iii) the lightest of the messenger fields is a heavy
stable particle which may lead to cosmological difficulties. Moreover, in these
models there is no apriori reason why the vector-like quarks cannot mix with the
known quarks. If they do mix, then additional tree level FCNC effects can spoil
the above naturalness property. Finally, these models do not address some other
generic problems of the MSSM such as the existence of R-parity breaking
interactions that lead to  arbitrary couplings for the unwanted baryon and
lepton number  violating couplings.

Since the general idea of the gauge mediated supersymmetry breaking is
attractive, it would be useful to explore models that avoid its undesirable
features while at the same time maintaining the good ones (such as the non
degeneracy of squark and slepton fields). With this goal in mind,  two of us
(S.N. and R.N.M.) began exploring a new class of models\cite{satya} where the
messenger sector consists of different fields. We also chose the electroweak
gauge group  to be
$SU(2)_L\times U(1)_{I_{3R}}\times U(1)_{B-L}$\cite{affleck} so that it
guarantees automatic R-parity conservation\cite{moh}. It then turns out that the
messenger fields can also play the role of Higgs fields that can serve to break 
the above gauge group down to the standard model group. Thus the scale of
supersymmetry breaking gets connected to the scale of gauge symmetry breaking
and the breaking of electroweak symmetry remains radiative.  This is a very
minimal extension of the standard model which also leads to nonzero neutrino
masses via the usual see-saw mechanism\cite{gell}. Moreover, since the messenger
fields couple to the right-handed neutrinos in order  to implement the see-saw
mechanism, they are unstable and therefore do not cause any cosmological problem.

It is the goal of this paper to study a wider class of models which follow the
spirit of this idea but use different messenger fields. We study the following
different messenger sectors: (the numbers in the parenthesis denote the
$SU(2)_L\times U(1)_{I_{3R}}\times U(1)_{B-L}$ quantum numbers) (AI) the minimal
one of Ref.\cite{satya} which uses only fields $\delta(1, 1, -2)+\bar{\delta}(1,
-1,+2)$; (AII) one pair of $\delta(1, 1, -2)+\bar{\delta}(1, -1, +2)$  along
with two extra Higgs  doublets $H'_u(2, 1/2, 0)$ and $H'_d(2, -1/2, 0)$; (AIII)
model (AII) with the addition of a pair of color octets (electroweak singlet)
$Q$. We also extend the gauge group to the left-right symmetric $SU(2)_L\times
SU(2)_R\times U(1)_{B-L}$ case. We impose the requirement of electroweak
symmetry breaking to be radiative and study the detailed predictions for the
sparticle spectra for these models and outline the phenomenological
implications. We find that the present experimental data seem to allow only the
third model for both the gauge groups. Once the requirement of radiative 
electroweak symmetry breaking is relaxed, all the cases become viable. We
present typical  particle spectra for each of these cases. They turn out to be
very  different from the conventional GMSB models making it  possible to test
the general idea of the gauge mediated supersymmetry   breaking as opposed to a
specific messenger version of it. A distinct and testable prediction of our
models is that the lightest neutralino is always the NLSP leading to photonic
events in the colliders.

We arrange this paper as follows: in section II, we outline the model and
present the mass formulae for the various sparticles at the scale $\Lambda$ of
supersymmetry breaking; in section III, we present the particle spectra for
these models and we discuss the phenomenological implications and tests of the
models. Section IV contains the minimization of the Higgs potential for one
typical version of these models and in section V,  we present some concluding
remarks. 
                   
\section{The Models}

\begin{center} {\bf A: Models with $SU(2)_L\times U(1)_{I_{3R}}\times
U(1)_{B-L}$}
\end{center}

We begin our discussion with a brief recap of the model of Ref.\cite{satya}
which we call model AI. The electroweak gauge group is $SU(2)_L\times
U(1)_{I_{3R}}
\times U(1)_{B-L}$ with quarks and leptons (including the right-handed
neutrinos, $\nu^c$) transforming as follows: $Q (2, 0, 1/3)~~$; 
$L (2, 0, -1)~~$;
$u^c (1, -1/2, -1/3)~~$; $d^c (1, +1/2, -1/3)~~$; $e^c (1, +1/2, +1)~~$; 
$\nu^c (1, -1/2, +1)$. The two MSSM Higgs doublet superfields transform as $H_u
(2, +1/2, 0)$ and $ H_d (2, -1/2, 0)$. In addition to these, we add the fields
$\delta $ and $\bar{\delta}$ (three such pairs)  that break the
$U(1)_{I_{3R}}\times U(1)_{B-L}$ symmetry down to the $U(1)_Y$ of the standard
model. They have the quantum numbers
$\delta (1, +1, -2)$ and $\bar{\delta} (1, -1, +2)$. The superpotential for the
matter sector of the theory (denoted $W_{vis}$ is given by:
                                                   
\begin{eqnarray} W_{vis} = h_u QH_u u^c + h_d Q H_d d^c + h_e L H_d e^c +
h_{\nu} L H_u \nu^c + \mu H_u H_d + f \delta \nu^c\nu^c.
\end{eqnarray} This is common to all versions of the type A models. The
different cases arise from the different choices of the messenger sector.

\noindent{\it Model AI}
                                     
The messenger sector in this case consists of three pairs of fields $\delta$ and
$\bar{\delta}$ mentioned above. The messenger sector superpotential $W_m$ 
(representing only a part of the complete superpotential) is given by 
\begin{eqnarray} W_m = \lambda S \delta \bar{\delta} + M_{\delta} \delta
\bar{\delta}.
\end{eqnarray}                                      
We will show in section. V that the 
$F_S$, $\delta$ and $\bar{\delta}$  acquire nonzero vacuum expectation values
(VEV) so that supersymmetry  as well as $U(1)_{B-L}\times U(1)_{I_{3R}}$ are
broken at the same scale.  As a result, the supersymmetry breaking scale and the
$B-L$ breaking scale get linked to each other and cannot be arbitrarily adjusted
in the physics discussion. 

 Next, we note that the supersymmetry breaking  is transmitted from the hidden
sector to the visible sector  via the $B-L$ and $I_{3R}$ gauge  interactions. As
a result, it is the exchange of
$B-L$ and $I_{3R}$ gauge particles that replaces the standard model (and MSSM)
particles in the GMSB model graphs  that give a Majorana mass to the gauginos at
the one loop and sfermions at the two loop level.  The resulting Majorana mass
of the $B-L$ and $I_{3R}$  gauginos ($\lambda_{B-L, I_{3R}}$) is such that the
Bino 
$\lambda_Y\equiv (g^{-1}_{B-L}
\lambda_{B-L} + g^{-1}_{R}\lambda_{I_{3R}})$ remains massless.  In other words,
in the language of the MSSM, $M_1=0$. Furthermore, the
$SU(2)_L$ gauginos also do not have any Majorana mass (i.e. $M_2=0$). The gluino
is also massless in this model as was already noted in
\cite{satya}.  It is worth pointing out that there are tiny contributions to all
gaugino masses in this model at the electroweak scale. For instance,     gluino
masses arise from the diagram in Fig. 1 and can be estimated to be 
\begin{eqnarray} M_{\tilde{G}}\simeq \frac{\alpha_s}{4\pi} \frac{m^2_t\mu cot
\beta}{M_{\tilde t}^2}
\end{eqnarray} leading to a mass of the order of a GeV or less. The same holds
for all the models discussed here. This loop induced mass becomes important if
there are no other larger contribution in the model and leads to the light
gluino scenario advocated in recent literature\cite{farrar}. These, as we will
see, have profound implications for phenomenology. 

 Turning now to the remaining sfermions, we find that:
\begin{eqnarray} M^2_{\tilde{F}}\simeq 2[x^2_{F} 
\left(\frac{\alpha_{B-L}}{4\pi}\right)^2 
\Lambda^2_S + y^2_{F} \left( \frac{\alpha_R}{4\pi}\right)^2\Lambda_S^2]
\end{eqnarray} where $x_F$ and $y_F$ denote the
$\sqrt{\frac{3}{2}}\frac{B-L}{2}$ and $I_{3R}$  values for the different
superfields $F$ (both matter as well as Higgs) and $\Lambda_S$ denotes  ratio
$<F_S>/<S>$ where $<S>$ denotes the VEV of S field. It is therefore clear that
the good FCNC properties of the usual GMSB are maintained in this class of
models. Furthermore the spectrum of squarks and sleptons here is very different
from that of the usual GMSB models, where messenger fields carry color. For
example, the sleptons are heavier than the squarks. In section III,  we present
the predictions for the various sfermion masses in this model.
              
\noindent{\it Model AII}
                                                   
The second class of models with the same gauge group, we consider, is
characterized by the following Higgs content that changes the character of the
messenger sector. In addition to one pair of $\delta$ fields as in model $AI$ ,
we include two pairs of $SU(2)_L\times U(1)_Y$ Higgs doublets $H^{a}_u$ and
$H^{a}_d$ (with a=1,2). The standard model Higgs doublets will arise out of
these Higgs fields and below the SUSY breaking scale there are only light
doublets. We will assume that the second pair (a=2) does not couple to the
quarks and leptons. We will show in section V that the minimization of the Higgs
potential will lead to VEV's for the $\delta$ fields which will induce
supersymmetry breaking $F_S\neq 0$. The relevant part of the hidden sector
superpotential $W_m$  has the form
\begin{eqnarray} W_H = \lambda S H^1_uH^1_d + M_H H^1_u H^1_d
\end{eqnarray} The $H^1_u$ and $H^1_d$ will play the role of messenger fields.
It is then clear that now $M_{1,2}$ are nonzero removing a major constraint in
the phenomenological analysis. The gluino is however still  massless. The masses
$M_k$ with $k=1,2$ denoting the $SU(2)$ and $U(1)$ gauge groups are given by:
\begin{eqnarray} M_1 =\frac{3}{5} \frac{\alpha_1}{4\pi} \Lambda_S \\ \nonumber
M_2 =\frac{\alpha_{2L}}{4\pi} \Lambda_S
\end{eqnarray} As far as the other sfermion masses go, we give the relevant
formulae below:
\begin{eqnarray} M^2_{\tilde{Q}}= m^2_{2L} +\frac{1}{24} m^2_{BL} \\ \nonumber 
M^2_{\tilde{q^c}}= m^2_R + \frac{1}{24} m^2_{BL} \\ \nonumber M^2_{\tilde{L}}=
m^2_{2L}+ \frac{3}{8} m^2_{BL} \\ \nonumber M^2_{\tilde{e^c}}= m^2_R +
\frac{3}{8} m^2_{BL} \\ \nonumber M^2_{H_u} = m^2_{2L} + m^2_R = M^2_{H_d} 
\end{eqnarray} where
\begin{eqnarray} m^2_{2L} =2\Lambda^2_S
\frac{3}{4}\left(\frac{\alpha_2}{4\pi}\right)^2\\
\nonumber m^2_{BL} =2 \Lambda^2_S \left(\frac{\alpha_{B-L}}{4\pi}\right)^2\\
\nonumber m^2_R = 2\Lambda^2_S
\frac{1}{4}\left(\frac{\alpha_{R}}{4\pi}\right)^2.
\\ \nonumber
\end{eqnarray}    

\noindent{\it Model AIII}

The last class of models we will consider will have some messengers with color
so that the gluinos can pick up mass. We do not consider colored fields with
quantum numbers identical with the quarks as in the usual GMSB models since they
could mix with the known quarks and generate new undesirable FCNC effects. As an
example we will consider a pair of color octet fields along with two pairs of
Higgs doublets as in AII.  We will show in sec. IV that our superpotential is
such that these color fields do not acquire VEV's and therefore there is no
danger of color breaking. We will now have all gauginos picking up Majorana
masses. The $U(1)$ and $SU(2)_L$ gaugino masses are same as in the model {\it
AII}. The gluino mass is given by:
\begin{eqnarray} M_{\tilde{g}}= 3\frac{\alpha_s}{4\pi} \Lambda_S
\end{eqnarray} As far as the other sfermion masses are concerned, the slepton
and the Higgs masses are same as in the model {\it AII}; the squark masses are
given as follows:
\begin{eqnarray} M^2_{\tilde{Q}} = m^2_{2L} + \frac{1}{24} m^2_{BL} + m^2_{3c}\\
\nonumber M^2_{\tilde{q^c}}= m^2_R + \frac{1}{24} m^2_{BL} + m^2_{3c}
\end{eqnarray} where $m^2_{3c}= 8 \Lambda^2_S
\left(\frac{\alpha_s}{4\pi}\right)^2$.

\begin{center} {\bf B: Models with $SU(2)_L\times SU(2)_R\times U(1)_{B-L}$
gauge group}
\end{center}

Let us now turn to the models with the left-right symmetric gauge group. The
quarks and leptons transform under the gauge group as:
$Q(2,1,1/3)$; $Q^c(1,2,-1/3)$; $L(2,1,-1)$ and $L^c(1,2,+1)$. We implement the
breaking of $SU(2)_R\times U(1)_{B-L}$ symmetry by means of the triplet Higgs
pair as in the usual left-right models\cite{senj}.   The messenger sectors will
essentially be the same as in case (A) except that the doublet pairs everywhere
will now be replaced by bi-doublets
$\phi_a(2,2,0)$. We will assume that the spectrum below the scale $\Lambda_S$ is
same as in MSSM; this will require that we enforce a doublet-doublet splitting
at that scale. One way to do this is to follow a recent suggestion\cite{rasin}
where the parameters of the superpotential involving the bidoublets are
finetuned. This fine tuning is essentially the same as the unresolved 
$\mu$ problem of the MSSM and we do not have anything more to say on this
question. In any case, as far as the contribution to the gaugino masses go, the
$M_{\tilde{g}}$ will remain the same as before for the various cases, and the
$M_1$ and $M_2$ are as follows:
\begin{eqnarray} Model~~ BI: M_1= M_2=0;\\ \nonumber Model~~ BII~~ and~~ BIII:
M_1= \frac{3}{5}
\frac{\alpha_1}{4\pi} \Lambda_S; M_2=  \frac{\alpha_{2L}}{4\pi} \Lambda_S.
\end{eqnarray}   The left-chiral squark and slepton mass contributions remain
the same as in the  case {\it A}. As far as the masses of $q^c$, $L^c$ are
concerned, in formula Eq. (7), we now have
$m^2_R~=~2\Lambda^2_S\frac{3}{4}\left(\frac{\alpha_R}{4\pi}\right)^2$

\section{Sparticle Spectroscopy}

In this section, we discuss the sparticle spectroscopy of the type A models. The
predictions for the type B models are same for the left-chiral sparticles but
only slightly different for the right chiral ones. We therefore do not discuss
the type B models in this section.

In giving the numerical predictions for the sfermion masses in this model, we
start by first requiring that the electroweak symmetry be radiatively broken. As
is well-known this requires that:
\begin{eqnarray}
\frac{M^2_Z}{2} = \frac{M^2_{H_d}-M^2_{H_u}tan^2\beta}{tan^2\beta -1} - \mu^2
\end{eqnarray} where all radiative corrections are absorbed into the parameters.
The reasons why it is nontrivial to satisfy this equation in predictive models
such as the GMSB models are the following: the Higgs masses are predicted  to be
positive at the scale $\Lambda_S$ and  to be proportional to various gauge
couplings $\alpha_{i}$ and
$\Lambda_S$ and the renormalization group evolution of the Higgs masses depends
on the value of the stop masses at $\Lambda_S$. So for instance, if in a model,
the $M^2_{H_u}(\Lambda_S)$ is too large and the stop masses are not large enough
at that scale, $M^2_{H_u}$ will remain positive at $M_Z$ and will not lead to
electroweak symmetry breaking. In one of the models to be discussed later this
is what happens. To see this in a qualitative manner, let us write down the
extrapolated value of the $M^2_{H_u}(m_{\tilde{t}})$ in the lowest order
approximation (neglecting running of squark masses and the effect of the gaugino
masses).

\begin{eqnarray} m^2_{H_u}(m_{\tilde{t}})\approx
m^2_{H_u}(\Lambda_S)-\frac{3\lambda^2_t}{8\pi^2}ln(\Lambda_S/m_{\tilde{t}})
\left(
m^2_{H_u}(\Lambda_S)+m^2_{\tilde{Q}}(\Lambda_S)+m^2_{\tilde{u^c}}(\Lambda_S)\right)
\end{eqnarray}

In the minimal model ({\it AI}), it is clear from Eq. (13) that electroweak
symmetry breaking depends on the relative values of 
$\alpha_{B-L}$ and $\alpha_{R}$. The masses in the above Eqn. are given by:
\begin{eqnarray} m^2_{H_u}(\Lambda_S)\simeq \frac{1}{2}
(\frac{\alpha_R}{4\pi})^2\Lambda^2_S\\ \nonumber m^2_{\tilde{t}}(\Lambda_S)\simeq
\frac{1}{12}\left(\frac{\alpha_{B-L}}{4\pi}\right)^2
\Lambda^2_S\\ \nonumber m^2_{\tilde{t^c}}(\Lambda_S)\simeq
\left(\frac{1}{12}(\frac{\alpha_{B-L}}{4\pi})^2
+\frac{1}{2}(\frac{\alpha_R}{4\pi})^2\right)\Lambda^2_S
\end{eqnarray} It is clear from Eq.(13) and Eq.(14) that $m^2_{H_u}\leq 0$ for
$r^2\geq
\left((1-\frac{6ln(\Lambda_S/m_{\tilde{t}})}{8\pi^2})/(ln
\frac{\Lambda_S/m_{\tilde{t}}}{8\pi^2})\right)$,  where
$r=\alpha_{B-L}/\alpha_R$.  We find that for
$r\geq 3$, the radiative symmetry breaking becomes possible and that we can
satisfy the above EWSB equation for
$\Lambda_S\simeq 50 -100$ TeV for reasonable values of the $\mu$ parameter. We
should mention that this way of determining the allowed values of $r$ is very
rough and we have determined $r$ numerically. In table 1, we list the
predictions for the various sparticle masses (where we have numerically solved
the renormalization group equations exactly in the one loop approximation) for
two different choices of the scale
$\Lambda_S$,
$tan \beta$, 
$r\equiv \frac{\alpha_{B-L}}{\alpha_R}$ and sign($\mu$). We emphasize that these
are the only four inputs into the model and all other masses are predictions
(including the $\mu^2$ parameter, which is chosen to satisfy the EWSB
constraint). 

A detailed study shows that this model predicts the light Higgs mass to be in the
range of 30 to 50 GeV. The Higgs production cross-section in $e^+e^-$ collision
is of course slightly lowered by the factor $sin (\alpha -\beta)$\cite{dawson}.
There are two important decay modes of the lightest Higgs boson: $h\to b\bar{b}$
and
$\chi_1\chi_1$; however, detailed study of the neutralino mass matrix in the
allowed parameter space seems to give $h\to b\bar{b}$ mode to be the dominant
one. It is therefore hard to reconcile with the present LEPII data. Increasing
the value of $\mu$ does not seem to be of any help, since it would decrease the
chargino mass.
 The lightest neutralino in this model is primarily a combination of the
gauginos and next to lightest neutralino is a combination of the  Higgsinos. The
next to lightest neutralino is also very light (less than 20 GeV in the two
illustrative scenarios given in the Table1). Since these  next to lightest
neutralinos decay hadronically ( each neutralino would decay into a anti-quark,
a quark and a gluino), they would contribute to the hadronic decay width of Z.
However the contribution to the Z decay width is 7-15 MeV for the scenarios in
the Table 1 and is allowed by the present data. The lighter chargino pair
production cross-section  in this model (for the scenarios given in the Table 1)
is little higher
$\sim 4$ pb for $\sqrt s$=172 GeV.

It is worth emphasizing that a crucial assumption that leads to the above
conclusions is that the electroweak symmetry breaking be radiative. If however,
we give up that assumption and only assume that the supersymmetry breaking be
transmitted via $B-L$ gauge interactions, then Higgs and  the chargino masses can
become much higher and no such conflict with present data occurs. In such a
model, electroweak symmetry breaking may, for example, arise from terms like
$S(H_uH_d-v^2)$ in the superpotential.

Turning to the model {\it AII}, we find that the model fails to give rise to
electro-weak symmetry breaking. Fig. 2 shows the extrapolation of the Higgs
masses from the scale $\Lambda_S$ down to the weak scale and the main feature to
notice is that $M^2_{H_u}$ (and of course
$M^2_{H_d}$) remains positive. The main reason for this is the small value for
the stop masses at the $\Lambda_S$ scale, since it is the stop mass that drives
the up-type Higgs mass negative in the renormalization group evolution. One can
see this in a qualitative manner as follows. Note that we have
\begin{eqnarray} m^2_{H_u}(m_{\tilde{t}})\approx m^2_{2L} +m^2_R -
\frac{3\lambda^2_t}{8\pi^2}(2m^2_{2L} +2m^2_R
+\frac{1}{12} m^2_{BL})ln\frac{\Lambda_S}{m_{\tilde{t}}}
\end{eqnarray} Defining $y\equiv \frac{\alpha_2}{\alpha_R}$, we get
\begin{eqnarray}
\frac{m^2_{H_u}}{\Lambda^2_S}\left(\frac{4\pi}{\alpha_R}\right)^2 =(\frac{1}{2}
+\frac{3}{2} y^2)(1-2x)-\frac{x}{6} r^2
\end{eqnarray} where $x\equiv \frac{3\lambda^2_t
ln(\Lambda_S/m_{\tilde{t}})}{8\pi^2}$ is roughly $0.25$. The condition for the
$m^2_{H_u}$ to turn negative at the weak scale is $r^2\geq 6+18 y^2$ under the
constraint 
$1.9 r = 0.6 ry +0.4 y $. Only for a very narrow range of values of
$y\approx 2.9-3.4$, these two constraints are satisfied. We consider this to be
extreme fine tuning. r is also large in that range which implies very large
value of $\alpha_{B-L}$. Outside this very narrow range, the typical value of
$m^2_{H_u}$ is shown in Fig.2.

Let us now turn to the model {\it AIII} which includes the color octet
messengers. In this model the stop masses  at the scale $\Lambda_S$ are enhanced
due to the octet contribution. As a result, the electro-weak symmetry breaking
condition is satisfied more easily. In Table 2, we give the prediction for the
sparticle masses as well as the Higgs masses for this model.  The squark masses
in this model are close to one TeV and are higher than the prediction of the
usual GMSB models. The chargino masses in this model are well allowed by the
experimental data and due to the color octet contribution the gluino masses are
$\sim 1$ TeV. An important feature of this model is the prediction of a
relatively light neutralino, $\chi_0$ in the mass range of 54 GeV or less.
$\chi_0$ decays to a photon and a gravitino. Thus a production of this
neutralino pair in electron positron collider will give rise to two photons plus
missing energy in the final state. Both the OPAL and the ALEPH collaboration at
LEPII have looked for this signal. From the non observation of this signal at
$\sqrt s=172$ GeV, they have established the following bounds:\\ OPAL
Collaboration: $\sigma<$ 0.41 pb (95$\%$C.L.)\cite{OPAL}.\\ ALEPH Collaboration:
$\sigma<$ 0.18 pb\cite{ALEPH}.\\ In Fig (3a), we plot the cross-sections for
$e^{+}e^{-}\rightarrow
\chi_0\chi_0$ as functions of the neutralino mass for the three LEPII energies,
$\sqrt s$=172, 182, 194 GeV. The cross-sections in our model is much lower than
those in the usual GMSB model \cite{gmsb2} because of the larger mass of the
lighter scalar electron. These curves are for $\tan\beta=$9 and $r=$4. With the
increase of the value of $r$, the cross-sections gets smaller, whereas the
dependence on $\tan\beta$ is very small. As r increases,
$\alpha_{B-L}$ increases. We restrict to $r\le 6$ to keep $\alpha_{B-L}$ in the
perturbative region (for r=6, $\alpha_{B-L}(\Lambda_s)=0.075$). In Fig (3b), we
show that the values of the cross-sections for $\tan\beta=9$ and r=6. In both
Fig (3a) and (3b), $\Lambda_s$ has been varied from 40 to 60 GeV (the
corresponding values for the squark masses lie between 0.9 to 1.5 TeV). We do
not increase
$\Lambda_s$ any further in order to keep SUSY a viable explanation for the
hierarchy problem. From Figs (3a) and (3b), we see that the current LEP II
bounds on the neutralino pair productions are satisfied. The important point is
to note that the smallest allowed values of the cross-sections are not too much
below the current experimental bounds. Thus, the photon signals in our model
could be within reach in the LEPII experiments in very near future. 

Now we discuss the consequences of this model at the Tevatron collider
($\sqrt{s}=1.8$ TeV). In table 2, our lightest neutralino mass varies from 29
to 40 GeV, while the lighter chargino mass varies from 94 to 126 GeV. At Tevatron
these could be produced in the following processes:
\begin{eqnarray} p\bar{p}\rightarrow \chi^+_i\chi^-_j, \chi^{\pm}\chi_a,
\chi_a\chi_b
\end{eqnarray} where $i,j$ run over 1,2 while $a,b$ run from $0$ to $3$. For the
mass range given in table 2, $\chi^+_1\chi^-_1,~
\chi^{\pm}\chi_0,~\chi^{\pm}\chi_1$ can be produced at Tevatron with significant
cross-section. The subsequent  decays of the neutralino to a photon and a
gravitino and the decay of a  chargino to a $\ell\nu\gamma\tilde{G}$ or
$q\bar{q}\gamma\tilde{G}$ will give rise to inclusive $\gamma\gamma+$ missing
$E_T$ final states. Recently, the authors of Ref.\cite{AKKMM} have studied these
signals in great detail, including detector simulation. They conclude that with
the assumption of gaugino unification, $m_{\chi^+}\leq 125$ GeV and 
$m_{\tilde{N_1}}\leq 70$ GeV can be excluded with $100~pb^{-1}$ of Tevatron data
(such masses give rise to about 10 two photon events plus missing energy
inclusive events). Without the assumption of gaugino unification (which is the
case in our model), they exclude $m_{\chi^+}\leq 100$ GeV for $m_{\chi_0}\geq 50$
GeV. (In this analysis, $m_{\chi_0}\geq 50$ GeV is needed for the
photons and missing energy to satisfy the detector cut,
$p^{\gamma}_T\geq 12 $ GeV, missing $E_T\geq 30$ GeV). Thus the mass range 
presented in Table 2 are not excluded by the current Tevatron data, but in the
interesting boundary of being tested with the complete analysis of the Tevatron
data and could be easily tested in the upgraded Tevatron.

Another interesting feature of this model is that the lightest $\chi_0$ is always
the NLSP.  This is to be contrasted with  the usual GMSB models, where the NLSP
can be either the lightest neutralino or the lighter stau depending on the
parameter space. In addition, in the usual GMSB models, $\chi_0$ need not be as
light, whereas in our model as argued before, $\chi_0$ is in the 50 GeV range or
lighter because its mass is tied to the squark masses. The lighter chargino in
our model can be around 130GeV. Each chargino decays into a W and a neutralino,
where the neutralino decays $100
\%$ into
$\gamma$ and gravitino and the W decays partly into a electron and a
anti-neutrino. The final state has
$e^{+}e^{-}\gamma\gamma$ plus missing energy. For other decays of the $W$ boson,
we would in general have $l^+_il^-_j\gamma\gamma$ plus missing energy or
$l^{\pm}_i+jets+\gamma\gamma$ plus missing energy or multijet +$\gamma\gamma$
plus missing energy.  Such a signal with two hard photons will be easily
detected and will have negligible SM background. The detailed predictions for
this phenomenon at collider energies is presently under investigation and will
be the subject of a forthcoming publication\cite{we}.

If two pairs of Higgs doublets, instead of one, contribute to the  squark and
gaugino masses, then the lightest neutralino mass and  the chargino mass  become
larger due to the larger values of
$M_1$ and $M_2$ at the Gauge mediated scale. The  values of $M_1$ and $M_2$ are
larger by a factor of 4 than the previous case.  The squark masses do not get
affected much due to this new contribution, since the color octet has a bigger
contribution. The slepton masses also become larger. We show some scenarios for
this model in Table 3. It is interesting to note that in this case the squark
masses can be lower than the previous model. This is because the lightest
neutralino mass is heavier in this model which  reduces the
$\Lambda$ and  subsequently reduces the squark masses.  The pseudo scalar mass
and the charged Higgs mass can also be lower in this case due to the same
reason. The superpartner masses in this case can be  safely beyond the present
lower limits.

The mass spectrum for the models BI-BIII are almost comparable to the model
AI-AIII. 
\section{An explicit model for the Hidden sector}         

In this section we address the question of the explicit model that leads to a
VEV for $F_S$  used in the previous section while at the same time allowing the
appropriate messengers to transmit the supersymmetry breaking. This discussion
is nontrivial for the following reasons. While it is easy to costruct a
superpotential that leads to a singlet having  a non-zero VEV and a non-zero
$<F_S>$, it is not  simple to communicate the supersymmetry breaking to the
visible sector. For some of the problems see the paper by Dasgupta et
al\cite{randall}. Below we provide an explicit superpotential which enables us
to attain all our goals simultaneously\cite{dobrescu}. Furthermore, we will not
need Fayet-Iliopoulos terms to break supersymmetry.

Let us illustrate our method using an example from the class {\it A} models with
$\delta$ and ${\bar{\delta}}$ and two pairs of Higgs doublets:
$H_{u,d}$ and $H'_{u,d}$. The generalization to the other cases is
straightforward. We choose the superpotential of the form:
\begin{eqnarray} W_H = \lambda S(\delta\bar{\delta} - M^2 + H_u H_d) + 
\lambda' S' \delta\bar{\delta} + M_1(H_uH'_d+ H_d H'_u) + M_2 H_uH_d.
\end{eqnarray} The potential can be written down from this as follows:
\begin{eqnarray} V&=& V_F + V_D \\ \nonumber V_F&=& \lambda^2
|\delta\bar{\delta}- M^2 +  H_uH_d|^2 +\lambda' |\delta\bar{\delta}|^2 +
|\lambda S+\lambda' S'|^2 (|\delta|^2+|\bar{\delta}|^2)\\ \nonumber &+&
M^2_1(|H_u|^2+|H_d|^2)+|\lambda SH_d +M_2H_d+M_1H'_{d}|^2 +|\lambda SH_u
+M_1H'_u+M_2H_u|^2
\end{eqnarray} The D-terms are not shown but they are given by standard
expressions. We choose $M_1\gg M, M_2$. It is then easy to see that global
minimum of the theory corresponds to $<H_{u,d}>=0$;
$<\delta>^2=<\bar{\delta}>^2 =\frac{\lambda^2M^2}{\lambda^2+\lambda'^2}$ and 
$F_S=\frac{\lambda\lambda'M^2}{\lambda^2+\lambda'^2}$. To get the gaugino
masses, we do not need the $S$ VEV since the mass term $M_2$ plays that role in
this theory. An important point to emphasize is that in the tree level
$\lambda S +\lambda' S'$ vanishes giving rise to a flat direction. This is
however stabilized giving small $S$ VEV once loop effects are taken into account.

The important point to note is that the fields $H_{u,d}$ which play the role  of
messenger fields do not have VEV. This was guaranteed by the fact that there is
a second pair of similar fields and that the mass parameter $M_1$ is larger than
other mass parameters in the theory.  One can therefore include any pair of
fields (including colored fields) as messenger fields and keep them from
acquiring VEV provided we add a second identical pair and add a mass term
analogous to $M_1$ which is large. Note that, below the SUSY breaking scale we
have only a pair of light Higgs doublets.

Let us now apply the above discussion to the model AI.  We start with three
pairs of
 $\delta, \bar{\delta}$ which have the same quantum numbers under the gauge
group $SU(2)_L\times U(1)_{I_{3R}}\times U(1)_{B-L}$ as before and distinguished
from each other by a prime. We then write the following superpotential:
\begin{eqnarray} W=\lambda S (\delta\bar{\delta} - M^2 +\delta''\bar{\delta}'') 
+\lambda' S'\delta\bar{\delta}+ M_1 (\delta''
\bar{\delta}'+
\delta '\bar{\delta}'') + M_2 \delta'' \bar{\delta}''
\end{eqnarray} It is easy to see that for $M_1\gg M, M_2$, the ground state
corresponds to
$F_S$, $F_{S'}$ and $<\delta>=<\bar{\delta}>$ having nonvanishing VEV's exactly
as in the case above and $<\delta''>=<\bar{\delta}''>=0$. The  supersymmetry
breaking is then transmitted via the $\delta$ fields to the visible sector.

An exactly analogous construction applies to the case AIII with two pairs of
color octets\cite{octets}  replacing the doublets in the case above. As far as
the left-right symmetric group is concerned, we replace the the $\delta$'s above
by the $SU(2)_R$ triplets and the doublets by the appropriate bidoublets. Again
the above discussion carries through in a straight forward manner.

Finally, we wish to note that, it is possible to give a superpotential for the
case AI, for which both the electroweak as well as the $U(1)_{B-L}$ symmetry
breaking arises from radiative corrections. We discuss this in Appendix A. The
phenomenological profile of this model is similar to the case AI with the
difference that there is an extra light gaugino (with mass in the 100 GeV
range). This model will also predict a light Higgs in the range of 30 to 40 GeV
and hence is not consistent with present observations. We present it in the
appendix in any case since it has the amusing feature that all gauge symmetries
in this model are broken radiatively and $B-L$ is the sole mediator of
supersymmetry breaking.

\section{Discussions and conclusion}

In this paper we have investigated the sparticle spectroscopy and tests of a new
class of gauge mediated supersymmetry breaking models with the gauge group
$SU(2)_L\times U(1)_{I_{3R}}\times U(1)_{B-L}$ and its left-right symmetric
version.The main reason for choosing the alternative messenger sectors is that
we want to maintain R-parity conservation and the good FCNC properties while
keeping the model phenomenologically viable. In one of the models, model AI, the
supersymmetry breaking is transmitted via the $B-L$ gauge interactions. In this
case however, we find that the light Higgs mass is too low for most of the
parameter range that we investigated. We then investigate alternative messenger
sectors involving color octets and $SU(2)_L$ doublets. We find that when we have
a pair of  color octets and a pair of
$SU(2)_L$ doublets, the model $SU(2)_L\times U(1)_{I_{3R}}\times U(1)_{B-L}$
has  the interesting prediction that the lightest neutralino is always the NLSP
and is very light (in the 50 GeV range or less). This would therefore be
testable at LEPII through the direct production of the neutralinos or at
Tevatron through the decays of the charginos that would eventually lead to
photonic events accompanied by lepton pairs, leptons and jets or pairs of jets.
 The lightest neutralino mass is however larger in another version of the model
where we have two pairs of doublets contributing to the SUSY breaking soft
masses. In this case the masses of the SUSY particles can be safely beyond the
present experimental bound. The left-right version of the model also has similar
results. 
\begin{center} {\bf Acknowledgement}
\end{center}

We are very thankful to D. Strom and G. Snow for useful comments regarding the
experimental aspects of this work. We also like  to acknowledge useful
discussions and communications  with C. Y. Chang, G. Farrar, E. Gross, A.
Jawahery, M. Luty and M. Schmitt. The work of B. D. is  supported by the DOE
grant no. DE-FG06-854ER-40224. The work of R. N. M. is supported  by the
National Science Foundation  grant no. PHY-9421386 and the work of S. N. is
supported by the DOE  under grant no. DE-FG02-94-ER40852.

\newpage

\begin{center} {\bf Appendix A}
\end{center}

In this appendix we discuss a superpotential where the scales of supersymmetry
breaking and the $B-L$ symmetry breaking are decoupled from each other and all
gauge symmetry breakings arise from the radiative renormalization group
evolution of the soft breakings. Furthermore the supersymmetry breaking is
transmitted to the visible sector only via the $B-L$ gauge interactions. The
phenomenological profile of this model is similar in all respects to the model
AI except that the $B-L$ gaugino in this model is also light. It therefore leads
to a light Higgs mass which is too low in the absence of any new contributions.
Nevertheless we present the model since it is particularly simple and could be
of interest as a realistic model if any new contribution to the Higgs mass could
be found.

The messenger sector of the model consists only of two pairs $\delta,
\bar{\delta}$ fields and a singlet field S. The superpotential of the model is
given by:
\begin{eqnarray} W= \lambda S (\delta\bar{\delta}-M^2) + M_1
(\delta\bar{\delta}' +\delta'\bar{\delta}) + M_2 \delta \bar{\delta}
\end{eqnarray}
 For $M_1\gg M, M_2$, the ground state of this theory has all $\delta$'s having
zero vevs and $F_S=-\lambda M^2$. Thus supersymmetry is broken at the scale and
can be transmitted to the visible sector via the $\delta$ fields. A combination
of the $B-L$ and $I_{3R}$ gaugino acquires mass at the one loop level via the
usual GMSB diagrams and the squarks and the sleptons acquire masses at the two
loop level. Once we include the
$f\nu^c\nu^c\delta$ term in the superpotential to implement the see-saw
mechanism, there appears another minimum of the potential with
$<\tilde{\nu^c}>, <\bar{\delta}'>\neq 0$ but other fields  with zero VEV which
is degenerate with the other minimum. This leads to the breakdown of
$U(1)_{I_{3R}}\times U(1)_{B-L}$ down to $U(1)_Y$.  This model then leads to
dynamical breaking of R-parity symmetry which however conserves baryon number.

\newpage

\noindent {\bf Table Caption:}

\noindent{\bf Table 1:} Mass spectrum for the superpartners in the scenarios 1
to 5 in Model A1(1 pair of Higgs doublet contribute to the soft supersymmetry
breaking masses);
 1st and 2nd generation superpartner masses are almost same.

\noindent{\bf Table 2:} Mass spectrum for the superpartners in the scenarios 1
to 5 in Model AIII(1 pair of Higgs doublet along with color octet contribute to
the soft supersymmetry breaking masses); 1st and 2nd generation superpartner
masses are almost same.

\noindent{\bf Table 3:} Mass spectrum for the superpartners in the scenarios 1
to 5 in Model AIII (2 pairs of Higgs doublets along with color octet contribute
to the soft supersymmetry breaking masses); 1st and 2nd generation superpartner
masses are almost same.

\noindent {\bf Figure Caption:}

\noindent {\bf Fig. 1}: One loop graph contributing to the gluino mass.

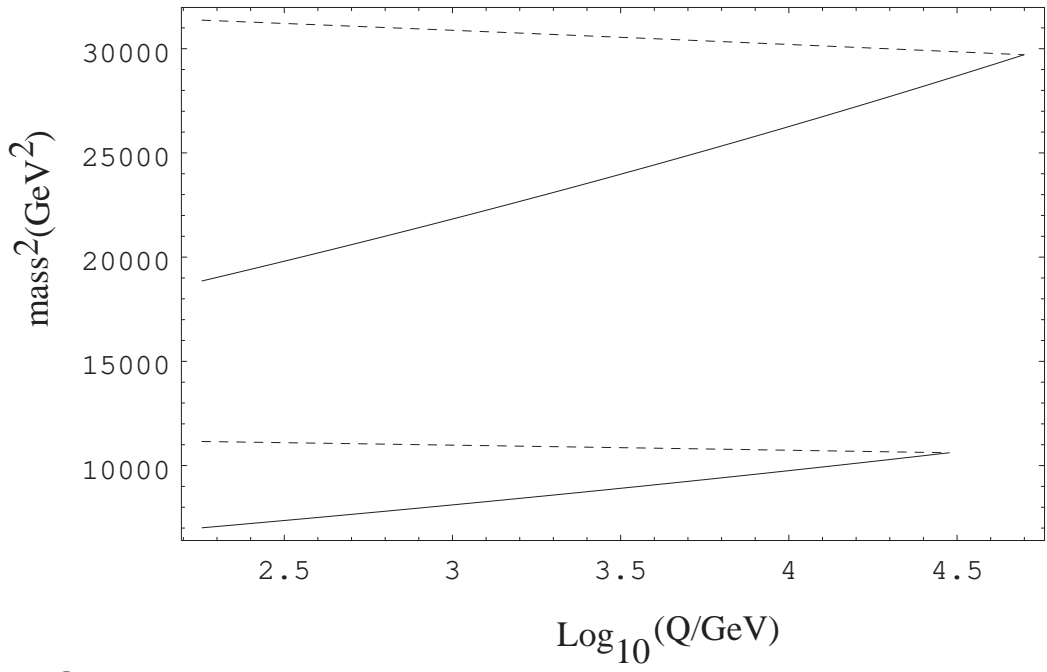
\noindent{\bf Fig. 2}: Running of the $m^2_{H_u}$ and $m^2_{H_d}$ in model AII.
The solid line corresponds to $m^2_{H_u}$ and the dashed line corresponds to
$m^2_{H_d}$. The pair of lines in the bottom of the figure corresponds to
$\Lambda_s=$ 50 TeV, $r(\equiv \alpha_{B-L}/\alpha_R)=3.33$ and $\tan\beta$=3
and the pair of lines in the top corresponds to
$\Lambda_s=$ 30 TeV, $r(\equiv \alpha_{B-L}/\alpha_R)=3.33$ and $\tan\beta$=3. 
                        
\noindent {\bf Fig. 3}: a) The value of the cross-section for $\sigma (e^+e^-
\to \chi^0\chi^0)$ as a function of the $\chi^0$ mass at various LEP energies
for $r(\equiv \alpha_{B-L}/\alpha_R)=4$ and $\tan\beta$=9. The dot-dashed line
corresponds to $\sqrt s=$ 194 GeV, the solid line corresponds to $\sqrt s=$ 182
GeV and the dashed line corresponds to $\sqrt s=$ 172 GeV.\\
\noindent b) The value of the croos-section for $\sigma (e^+e^-
\to \chi^0\chi^0)$ as a function of the $\chi^0$ mass at various LEP energies
for $r=6$ and $\tan\beta$=9.

\newpage
\begin{center}  Table 1 \end{center}
\begin{center}
\begin{tabular}{|c|c|c|c|}  \hline &Scenario 1&Scenario 2
\\\hline masses&$\Lambda=79$ TeV,&$\Lambda=75.5$ TeV,\\ 
&$\tan\beta$=1.8,&$\tan\beta$=1.9,\\ &r=3&r=3.33\\\hline 
m$_h$(GeV)&33&32\\\hline m$_{H^{\pm}}$&96&91\\\hline  m$_A$&52&43\\\hline
m$_{\chi^0}$&1&1\\\hline  m$_{\chi^1}$&14&14\\\hline m$_{\chi^2}$&84&84\\\hline 
m$_{\chi^3}$&99&100\\\hline m$_{\chi^{\pm}}$&54,101&51,103\\\hline  m$_{\tilde
{\tau}_{1,2}}$&284,290&296,302\\\hline  m$_{\tilde
{e}_{1,2}}$&284,291&296,302\\\hline  m$_{\tilde {\nu}_{L}}$&282&294\\\hline 
m$_{\tilde {\rm t}_{1,2}}$&86,200&89,200\\\hline  m$_{\tilde {\rm
b}_{1,2}}$&94,115&98,116\\\hline  m$_{\tilde {\rm
u}_{1,2}}$&86,110&111,189\\\hline  m$_{\tilde {\rm
d}_{1,2}}$&94,115&107,116\\\hline  m$_{\tilde g}$&0.4&0.4\\\hline
$\mu$&-19&-18\\\hline
\end{tabular}
\end{center}
\newpage
                   
\begin{center}  Table 2 \end{center}
\begin{center}
\begin{tabular}{|c|c|c|c|c|c|c|}  \hline &Scenario 1&Scenario 2&Scenario 3
&Scenario 4&Scenario5
\\\hline masses&$\Lambda=40$ TeV,&$\Lambda=50$ TeV,&$\Lambda=45$ TeV,
&$\Lambda=43$ TeV,&$\Lambda=47$ TeV,\\ 
&$\tan\beta$=3,&$\tan\beta$=9,&$\tan\beta$=15,&$\tan\beta$=25&$\tan\beta$=30,\\
&r=4&r=3.3&r=3.3&r=3.3&r=3\\\hline  m$_h$(GeV)&113&128&127&127&129\\\hline
m$_{H^{\pm}}$&495&539&466&420&446\\\hline  m$_A$&489&532&459&412&438\\\hline
m$_{\chi^0}$&29&40&36&34&38\\\hline  m$_{\chi^1}$&95&127&114&109&123\\\hline
m$_{\chi^2}$&447&512&452&430&474\\\hline 
m$_{\chi^3}$&462&519&459&436&479\\\hline
m$_{\chi^{\pm}}$&94,463&126,522&114,467&109,441&120,484\\\hline  m$_{\tilde
{\tau}_{1,2}}$&185,230&196,267&167,247&138,249&123,266\\\hline  m$_{\tilde
{e}_{1,2}}$&186,231&202,262&182,247&174,227&176,237\\\hline m$_{\tilde
{\nu}_{L}}$&226&259&233&222&232\\\hline   m$_{\tilde {\rm
t}_{1,2}}$&841,923&1067,1140&964,1026&922,979&1006,1064\\\hline  m$_{\tilde {\rm
b}_{1,2}}$&900,924&1126,1156&1011,1040&958,994&1041,1082\\\hline  m$_{\tilde {\rm
u}_{1,2}}$&925,933&1156,1167&1040,1050&993,1003&1086,1097\\\hline  m$_{\tilde
{\rm d}_{1,2}}$&926,937&1157,1170&1041,1053&994,1007&1087,1100\\\hline 
m$_{\tilde g}$&979&1224&1101&1053&1151\\\hline
$\mu$&-446&-506&-447&-423&-451\\\hline
\end{tabular}
\end{center}
\newpage
\begin{center}  Table 3 \end{center}
\begin{center}
\begin{tabular}{|c|c|c|c|c|c|c|}  \hline &Scenario 1&Scenario 2&Scenario 3
&Scenario 4&Scenario5
\\\hline masses&$\Lambda=28$ TeV,&$\Lambda=35$ TeV,&$\Lambda=45$ TeV,
&$\Lambda=30$ TeV,&$\Lambda=47$ TeV,\\ 
&$\tan\beta$=3,&$\tan\beta$=9,&$\tan\beta$=15,&$\tan\beta$=25&$\tan\beta$=30,\\
&r=4&r=3.3&r=3.3&r=3.3&r=3\\\hline  m$_h$(GeV)&107&122&127&121&129\\\hline
m$_{H^{\pm}}$&349&380&488&293&464\\\hline  m$_A$&338&370&478&279&451\\\hline
m$_{\chi^0}$&82&109&144&92&150\\\hline  m$_{\chi^1}$&207&247&339&198&357\\\hline
m$_{\chi^2}$&253&274&364&224&380\\\hline 
m$_{\chi^3}$&353&4061&507&351&526\\\hline
m$_{\chi^{\pm}}$&201,352&244,406&339,508&319,485&358,527\\\hline  m$_{\tilde
{\tau}_{1,2}}$&256,327&277,376&354,485&231,324&332,486\\\hline  m$_{\tilde
{e}_{1,2}}$&256,327&278,376&357,485&238,322&344,486\\\hline  m$_{\tilde {\rm
t}_{1,2}}$&601,676&758,831&968,1062&655,713&1009,1100\\\hline  m$_{\tilde {\rm
b}_{1,2}}$&652,659&812,825&685,727&983,1012&1068,1102\\\hline  m$_{\tilde {\rm
u}_{1,2}}$&651,680&812,849&1045,1093&696,727&1091,1141\\\hline  m$_{\tilde {\rm
d}_{1,2}}$&653,683&815,853&698,731&1001,1047&1093,1144\\\hline  m$_{\tilde
g}$&685&857&1101&734&1151\\\hline
$\mu$&-250&-268&-360&-216&-378\\\hline
\end{tabular}
\end{center}

\newpage
\begin{figure}[htb]
\centerline{ \DESepsf(cdmnfig1.epsf width 15 cm) } \smallskip
\nonumber
\caption{}
\end{figure}

\newpage
\begin{figure}[htb]
\centerline{ \DESepsf(cdmnfig23.epsf width 15 cm) } \smallskip
\nonumber
\end{figure}

\newpage
\begin{figure}[htb]
\centerline{ \DESepsf(cdmnfig3.epsf width 15 cm) } \smallskip
\nonumber
\end{figure}
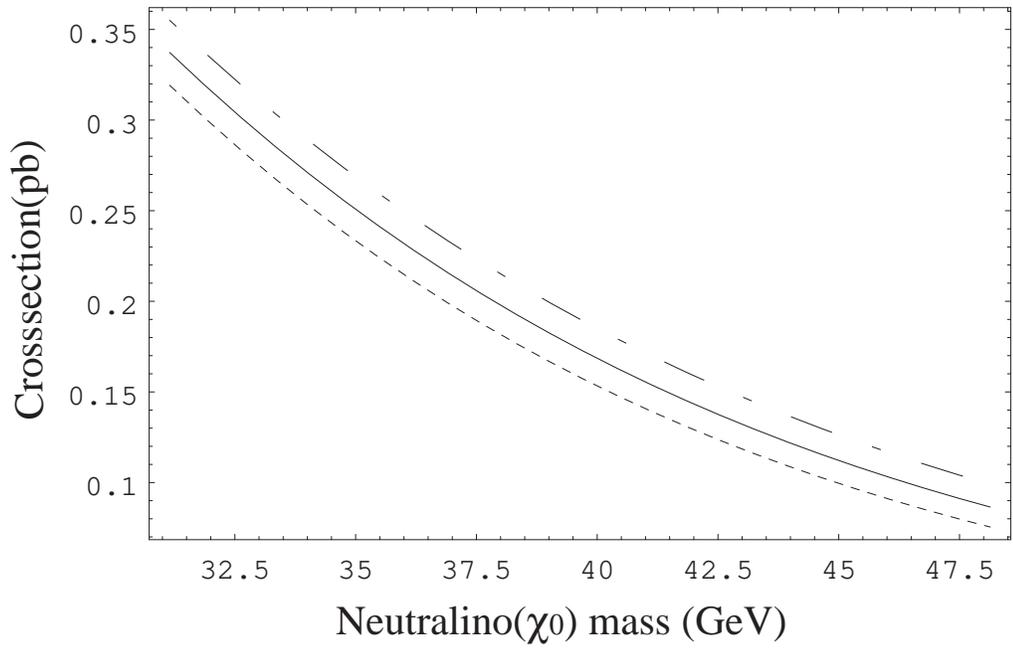
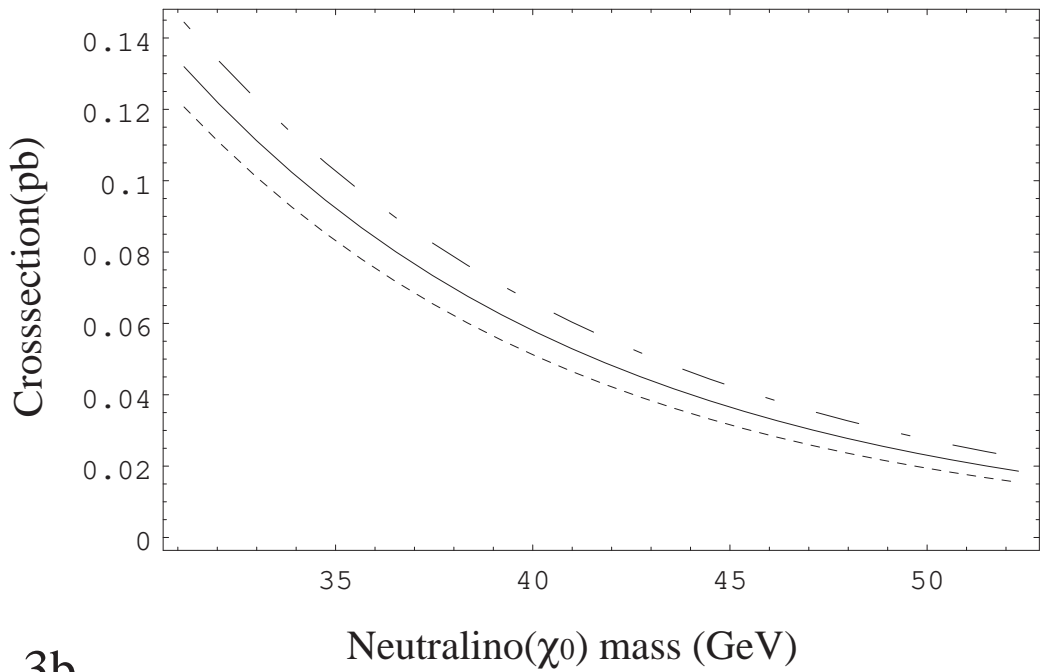
\end{document}